\documentstyle[12pt]{article}
\begin{document}
\thispagestyle{empty}
\noindent
\begin{flushright}
        OHSTPY-HEP-T-97-002\\
        February 1997
\end{flushright}

\vspace{1cm}
\begin{center}
  \begin{Large}
  \begin{bf}
      Gauge-mediated SUSY Breaking at an Intermediate Scale
   \\
  \end{bf}
 \end{Large}
\end{center}
  \vspace{1cm}
 
    \begin{center}
    Stuart Raby\\
      \vspace{0.3cm}
\begin{it}
Department of Physics,
The Ohio State University,
Columbus, Ohio  43210\\
\end{it}
  \end{center}
  \vspace{1cm}
\centerline{\bf Abstract}
\begin{quotation}
\noindent
Gauge-mediated SUSY breaking with a messenger scale of order $10^{15}$ GeV has 
some interesting features.   It can solve the flavor changing neutral current 
problem of supersymmetric models with predictions for superpartner masses 
which are identical to those of minimal supergravity models.  It can however 
also lead to theories with novel experimental signatures.  For example, we 
present a  model in which the gluino is the lightest supersymmetric particle.
We also review the present experimental status of a stable gluino.
 
\end{quotation}
\vfill\eject

\section{Introduction}
Supersymmetry [SUSY] breaking in a ``hidden" sector can be transmitted to the 
observable sector through Standard Model [SM] gauge 
interactions\cite{lsgauge,hsgauge} or through supergravity 
interactions\cite{sugra}.    In supergravity theories,  SUSY breaking in
the observable sector is set by the gravitino mass with $$m_{3/2} = {F \over 
\sqrt{3} M_{Planck}} .$$ $F$ is the fundamental scale of SUSY breaking and 
$$M_{Planck} \equiv 1/\sqrt{8 \pi G_N} = 2.4 \times 10^{18} GeV$$ is the 
messenger scale.   In gauge mediated SUSY breaking, on the other hand, $M$ is 
the mass of the messenger states which have the property that they
\begin{itemize} \item couple  directly to the SUSY breaking sector, and 
\item communicate with the observable sector predominantly through SM gauge 
interactions.\footnote{Gravity mediated SUSY breaking is always present.} 
\end{itemize}    Low energy messenger models have been discussed 
recently\cite{dn}.  In these theories, soft SUSY breaking mass terms, for 
scalars  and gauginos, scale as local dimension 2 and 3 operators, respectively, 
up to the messenger scale.  Above $M$, soft SUSY breaking mass terms no 
longer scale as local mass operators (see Appendix 1). 
As a result, low energy messenger models solve the flavor changing neutral 
current [FCNC] problems of SUSY\cite{dg,hkr}, since 
\begin{itemize} 
\item scalars with the same SM quantum numbers are degenerate, and 
\item the low energy physics is not sensitive to the physics above $M$.
\end{itemize}   In addition, these theories can lead  to observable low 
energy effects of SUSY breaking, i.e. a distinct spectrum of squark and 
slepton masses or new types of SUSY signatures for accelerator
experiments\cite{ddrt}.  

In the minimal messenger scenario\cite{dn}, scalars and gauginos get mass
of order 
\begin{equation}  {\alpha \over 4 \pi} {F_X \over M} \sim
{\alpha \over 4 \pi} M \end{equation} where  $F_X \sim M^2$, 
$M \sim 10 - 100 \; TeV$ and $\alpha$ is the fine structure constant associated 
with the SM charge carried by the particular scalar or gaugino.  In this paper 
we describe a simple generalization of the minimal messenger scenario which 
illustrates the following points --- 
\begin{enumerate}
 \item The SUSY breaking scale $\sqrt{F_X}$ and the messenger scale $M$ are not
necessarily the same.   The only necessity is that 
\begin{equation} \Lambda = {F_X \over M} \sim 10 - 100 \; TeV \end{equation} 
is fixed.  In our example we shall show that $M$ can have any value from 
$\sim 10 - 100 \; TeV$ to $M_{Planck}$.  Hence the two SUSY breaking 
schemes --  supergravity mediated SUSY breaking (with $M \sim M_{Planck}$) 
and the minimal gauge mediated SUSY breaking (with $M \sim 10 - 100 \; TeV$) 
are just two extreme limits in a continuum of possible theories. 

A particularly interesting case is when $M \approx M_{GUT}$\cite{hsgauge}. 
In this case, SUSY breaking is transmitted to the observable sector through 
GUT interactions. As a result, squarks and sleptons within a complete GUT 
multiplet are degenerate, {\em assuming} the messenger masses are invariant 
under the GUT symmetry.   Such theories look very much like the minimal 
supergravity models\cite{sugra} (or constrained minimal supersymmetric 
standard model[CMSSM]\cite{kane}).

\item In order to generate gaugino masses, it is necessary to break both SUSY 
and R symmetry.  In the minimal scenario, they are broken at the same scale.  
We shall give a simple example in which the R symmetry breaking scale $M_R$ 
is less than the messenger scale; thus suppressing gaugino masses.   

\item Finally, even in a GUT it is possible to obtain non-degenerate gaugino
masses at $M \approx M_{GUT}$.   In our example, gluinos may be the lightest
supersymmetric particles; resulting in very interesting experimental 
consequences.
\end{enumerate}

\section{Intermediate Messenger Scales}
 
We illustrate these ideas in a simple SO(10) SUSY GUT.\footnote{This model is 
not  ``natural" in the sense that it does not include all interactions not 
forbidden by symmetries.  It is however ``SUSY natural," since no new terms in 
the SUSY potential can be generated due to the non-renormalization theorems of 
SUSY.  However an example of a ``natural" theory of this type is given in 
Appendix 2.}  The superpotential is given by
\begin{equation} W = \lambda_{\cal S} {\cal S} 10^2 + X( 10^2 + \lambda 
\phi^+ \phi^- + \lambda' X^2) \label{eq:W} \end{equation}
where ${\cal S}, X$ are SO(10) singlets, $\phi^+, \phi^-$ are also SO(10)
singlets with charge $\pm 1$ under a messenger U(1) and 10 is a ten under 
SO(10). Note that we assume SUSY breaking occurs in a ``hidden" sector 
dynamically and transmits this directly to the states $\phi^+, \phi^-$ via 
the messenger U(1) resulting in a negative mass squared for these two 
states\cite{dn}.

The scalar potential is thus given by
\begin{eqnarray} 
V  = & \lambda_{\cal S}^2 |10^2|^2 + |10^2 + \lambda \phi^+ \phi^- + 3 \lambda' 
X^2|^2 &   \\  & + |(\lambda_{\cal S}{\cal S} + X) 10|^2 + \lambda^2 |X|^2 (|\phi^+|^2
+ |\phi^-|^2) & \nonumber \\ &  - m^2 (|\phi^+|^2 + |\phi^-|^2) + {1 \over 2} D_1^2 +
{1 \over 2} D_{10}^2 & \nonumber \end{eqnarray}
where $D_1 = g_1 (|\phi^+|^2 - |\phi^-|^2 + \cdots)$ and 
$D_{10} = g_{10}( |10|^2 + \cdots)$ and the ellipsis refers to the 
contribution to $D_1$($D_{10}$) of all other U(1)(SO(10)) multiplets necessary 
to make this a complete theory.

A minimum is obtained for vacuum expectation values [vev] $<\phi^+> = -<\phi^->
\equiv \phi, \; <X> \equiv X$  and $ <10> = 0$ with
\begin{equation}  \phi^2 = {3 \lambda' m^2 \over \lambda^3}  {1 \over (2 - 
\lambda/3 \lambda')}, \;\; X^2 = {m^2 \over \lambda^2} \left({1 - \lambda/3
\lambda' \over 2 - \lambda/ 3 \lambda'}\right)  \end{equation}
and $\lambda/3 \lambda' < 1$.  The vev $<{\cal S}>
\equiv M $ is undetermined in the global SUSY limit, at tree level.

At the minimum, SUSY is broken with 
\begin{equation} F_X = {m^2 \over \lambda} {1 \over (2 - \lambda/3 \lambda')}
\end{equation}  and
\begin{equation} V_{tree} = F_X^2 (1 - 6 \lambda' /\lambda)  < 0 . 
\end{equation}
Note,  $\sqrt{F_X} \sim X \sim m$;  a property of the minimal model.

We now argue that $M$ is the messenger scale when $M >> m$ and in section 2.2 we
show how the condition $M >> m$ can be obtained without any fine-tuning.   The
$10$ is the only field which couples to SO(10) and feels SUSY breaking directly
through $F_X$.   The mass of the $10$ is given by $\lambda_{\cal S} M + X 
\approx {\rm Max}(M,m) $.  Note, if $M \approx m$ then we have the light messenger
scenario. If on the other hand $M >> m$, then  the effective supersymmetry 
breaking scale is given by $\Lambda = F_X/M \approx m^2/M  \sim 10 - 100 \; TeV$
with both $M$ and $m$ much greater than $\Lambda$.
  
For $M_{GUT} >> M$, gauginos, squarks and sleptons receive mass 
radiatively, predominantly through SM gauge interactions\cite{lsgauge,dn,ddrt}.
Gauginos obtain mass at one loop given by
\begin{eqnarray}
    m_{\lambda_i} = & c_i  {\alpha_i(M)\over4\pi} \Lambda & ( {\rm for}\;\; i =
1,2,3) \nonumber 
\end{eqnarray}
where $c_1 = {5 \over 3},\; c_2 = c_3 = 1$. 

The scalar masses squared arise at two-loops
 \begin{eqnarray} \tilde m^2 = & 2 \Lambda^2 \left[
\sum_{i=1}^3 C_i\left({\alpha_i(M) \over 4 \pi}\right)^2\right] & 
 \label{eq:mtilde} \end{eqnarray}
where $C_3 = {4 \over 3}$ for color triplets and zero for singlets, 
$C_2= {3 \over 4}$ for weak doublets and zero for singlets, and $C_1 = 
{5 \over 3}{\left(Y\over2\right)^2}$, with the ordinary hypercharge $Y$
normalized as $Q = T_3 + {1 \over 2} Y$.   In the limit $M << M_{GUT}$, 
we have $\alpha_3(M) >> \alpha_2(M) > \alpha_1(M)$. Thus squark doublets and 
singlets are approximately degenerate, while right-handed sleptons are 
expected to be the lightest SUSY partners of SM fermions.  

Now consider the case $M \sim M_{GUT}$. In this case all SO(10) gauge bosons 
contribute equally to gaugino, squark and slepton masses.   For gauginos we 
obtain a common mass at  one loop   
\begin{eqnarray}
m_{\lambda} = & {\alpha_{GUT}\over 4 \pi} \Lambda &
\end{eqnarray}
 and for squarks and sleptons a positive mass squared at two loops  
\begin{eqnarray}
\tilde m^2 = & 2 \Lambda^2 \left[ C_{10} \left({\alpha_{GUT}\over 4 \pi}
\right)^2\right] & \end{eqnarray} 
 with $C_{10} = {45 \over 8}$ for 16s of SO(10).    Hence, 
in this case, squarks and sleptons are degenerate, thus satisfying the
boundary conditions usually assumed for minimal supergravity mediated SUSY 
breaking or the CMSSM.  

Note that the gaugino masses are the same order as the squark and slepton 
masses.\footnote{Thus two of the independent parameters of minimal
supergravity, the scalar mass parameter $m_0$ and the
gaugino mass $M_{1/2}$ of minimal supergravity, are related.} This is because
$W$ [ eqn. \ref{eq:W}] has an R symmetry in which all  fields have R charge 1 
and $W$ has  R  charge 3.  This R symmetry must be broken in order to generate 
a gaugino mass.  However the R symmetry breaking scale, 
$M_R = M$, is the same as the messenger scale; hence gaugino masses are not
suppressed in this model.  In the next section we give an example 
where the R symmetry breaking scale is much smaller than the messenger scale.

\subsection{Supergravity effects}

What about supergravity effects?  For the range of parameters satisfying  
$\sqrt{3} M_{Planck} >> 4 \pi M/\alpha$, the dominant contribution to 
observable scalar masses is gauge mediated.\footnote{Supergravity will induce 
small contributions to scalar masses given by  $m_0 \approx m_{3/2} =
 F/ \sqrt{3} M_{Planck}$.  It should be stressed that $F_X$ is not necessarily
the intrinsic supersymmetry  breaking scale, $F$,  since the gauge singlet field
may not be coupled directly  to the supersymmetry breaking sector.  For example,
in the model of Dine, Nelson , Nir and Shirman\cite{dn},  $F \gg F_X$.  However,
it is also perfectly possible that $F\sim F_X$\cite{kenscott}.  While $F_X$ 
determines the  superpartner masses, it is $F$ which determines the gravitino
mass.}  In the  case  $F \sim F_X$,  $M \sim M_{GUT} \approx 2 \times 10^{16}
GeV$, and $\alpha_{GUT}^{-1} \approx 24$,  the contribution of supergravity
mediated SUSY  breaking is comparable to that of gauge mediated SUSY breaking. 
This  supergravity contribution to squark and slepton masses could lead to
observable  FCNC processes, if it does not respect flavor symmetry.  In addition
both  contributions receive radiative corrections from physics between $M_{GUT}$
and $M_{Planck}$ which could also induce large flavor violating
effects\cite{hkr}.   However both of these flavor violating contributions can be
suppressed by  reducing $M$.   For example, with $M \sim M_{GUT}/10$, the
supergravity contribution to flavor violating scalar masses $\delta m^2/m^2$ is
reduced to a  few percent.\footnote{If $\Lambda = F/M$ is fixed, then $m_0^2 =
(F/\sqrt{3} M_{Planck})^2 = \Lambda^2 (M/\sqrt{3} M_{Planck})^2.$}  The flavor
violating  GUT corrections to the gauge mediated scalar masses are also reduced
by a  similar factor $(M/M_{GUT})^2$ (see Appendix 1).   Thus both contributions
to  the real part of $\delta m^2/m^2$ are reduced to a few percent which appears 
to be safe\cite{ggms}. Finally, for $M \sim M_{GUT}/10$, squarks and sleptons 
are {\em almost} degenerate and thus approximately satisfy the boundary 
conditions usually assumed for the CMSSM\cite{sugra,kane}.

\subsection{Why is $M >> m$?}

How do we obtain $M >> m$?  The effective potential for ${\cal S}$ is flat at 
tree level in the global SUSY limit.  It receives significant contributions 
both from supergravity and at one loop given by
\begin{equation}  V({\cal S}) = V_{tree} + m_{\cal S}^2(\mu) {\cal S}^2 + 16 
F_X^2ln[({\cal S} + X)/m]  \end{equation}
where $m_{\cal S}^2(\mu)$ is the supergravity generated soft SUSY breaking 
mass term evaluated at the scale $\mu = {\cal S}$.    $m_{\cal S}^2(\mu)$, 
$m_{10}^2(\mu)$, $m_{\lambda_i}(\mu)$, $\alpha_i(\mu)$, $\lambda_{\cal
S}(\mu)$, etc. satisfy coupled RG equations schematically given by
\begin{eqnarray}  {d m_{\cal S}^2 \over dt} = & {\lambda_{\cal S}^2 \over 4\pi}
(20 m_{\cal S}^2 + 40 m_{10}^2 +  \cdots ) &  \\
{d m_{10}^2 \over dt} = & {\lambda_{\cal S}^2 \over 4\pi}
(4 m_{\cal S}^2 + 8 m_{10}^2 +  \cdots) &- 4 \alpha_i C_i m_{\lambda_i}^2 
\nonumber \\
{d \over dt}\left( {m_{\lambda_i} \over \alpha_i} \right) = & 0 & \nonumber
\end{eqnarray} 
where the ellipsis refers to A terms which may or may not be significant,
$t \equiv ln(\mu/M_{Planck})/2 \pi$, $C_i$ are defined after eqn.
(\ref{eq:mtilde}) and we have not written the other equations.
Note that the RG equations for $m_{10}$ have both gauge and Yukawa contributions 
whereas for $m_{\cal S}$ only Yukawa contributions appear.  Moreover
$m_{\lambda_3}$ and $\alpha_3$ grow as $\mu$ decreases.  As a result we expect
that $m_{10}$ will vary slowly with scale, while $m_{\cal S}$, on the  other
hand, will vary significantly.  The Yukawa interactions tend to drive  
$m_{\cal S}^2$ to negative values.  For example the solution to the much 
simplified equation  \begin{equation} {d m_{\cal S}^2 \over dt} =  
{\lambda_{\cal S}^2 \over 4\pi} (40 m_{10}^2 ) = constant \end{equation} 
is given by
\begin{equation} m_{\cal S}^2(t) =  m_{\cal S}^2(0) + {\lambda_{\cal S}^2 
\over 8\pi^2}(40 m_{10}^2 ) ln(\mu/M_{Planck}) \end{equation}
If we now let $m_{\cal S}^2(0)= m_{10}^2 =  m_0^2$,\footnote{We have only 
included the supergravity contribution to scalar masses in the effective theory.} we find
\begin{equation} m_{\cal S}^2(t) =  m_0^2 \left(1 + 40 {\lambda_{\cal S}^2 \over 8\pi^2} 
ln(\mu/M_{Planck})\right) \end{equation} which goes through zero at
$\mu = M_{Planck} exp[{-8 \pi^2 \over 40 \lambda_{\cal S}^2}] $.  Since this 
mass term gives the dominant contribution to the effective potential for 
${\cal S}$, it is clear that the value of $M = <{\cal S}> \approx M_{Planck} 
exp[{- \pi^2 \over 5 \lambda_{\cal S}^2}]$ can take on any value between
$M_Z$ and $M_{Planck}$ without any fine-tuning.

\section{Light Gluinos}

In the above case the mass of the $10$ (the messenger mass $M$) is SO(10) 
invariant. We now discuss a simple variation of the model in which the 
messenger mass breaks SO(10).   Consider the same superpotential as in 
eqn. (\ref{eq:W}) with the addition of a term
\begin{equation}   10 A 10_H  \end{equation}
where $A, \; 10_H$ are an additional adjoint and ten dimensional representation 
of SO(10).   We presume that $A$ is also included in an SO(10) breaking sector 
of the theory and obtains a vev of order $M_{GUT}$ in the B-L direction.  
As a result the effective $10 - 10_H$ mass term takes the form
\begin{equation}{1 \over 2} (10 \;\; 10_H) (\begin{array}{cc} 2{\cal S} & A \\
                                                             A & 0
                                                             \end{array})
                                           (\begin{array}{c}  10 \\ 10_H
                                           \end{array}) \end{equation}
In this case the color triplets in $10 - 10_H$ obtain mass of order $M_{GUT}$, 
while the doublets in $10, \;\; 10_H$ obtain mass of order $M, \;\;0$ 
respectively\cite{dw}.   The massless doublets may be identified as the MSSM 
electroweak higgs doublets.  

Now if $M_{GUT} >> M >> m$ we must carefully identify both the messenger scale, 
and the R symmetry breaking scale, $M_R$; the SUSY breaking scale is still $m$.
First consider the R symmetry breaking scale.  The vev of $A$ appears to break
R symmetry and in fact it does break the R symmetry defined earlier.   However 
it also breaks a global U(1) symmetry in which $10,\; 10_H,\; A,\; {\cal S}$ 
have charge $0, -1, 1, 0$ respectively.  Thus there is an R symmetry with 
charges $1, 2, 0, 1$ which remains conserved.  This R symmetry is sufficient to 
keep gauginos massless.   The vev $<{\cal S}> = M$ on the other hand, breaks 
this R symmetry; hence $M_R = M$.   

There are two effective messenger scales in this model given by $M_{GUT}$ and 
$M$.  In the color non-singlet sector, the messenger scale is fixed by the vev 
$<A> = M_{GUT}\times (B-L)$, while in the color singlet sector the messenger 
scale is determined by the vev $<{\cal S}> = M$.  Thus there are now two 
effective SUSY breaking scales,  $\Lambda_C = F_X/M_{GUT}$ in the color 
non-singlet sector and $\Lambda = F_X/M$ for color singlets. Note $\Lambda 
> \Lambda_C$ for $M < M_{GUT}$. We take $\Lambda \sim 10 - 100 \; TeV$.

Integrating out the messenger sector in this case gives rise to gaugino 
masses at one loop.   We find
\begin{eqnarray}
   m_{\lambda_3} = &  c_3 {\alpha_3(M_{GUT})\over 4\pi} \; \Lambda_C \;({M \over 
   M_{GUT}}) & \\  m_{\lambda_i} = & c_i  {\alpha_i(M)\over4\pi} \; \Lambda & 
   ( {\rm for}\;\; i = 1,2)  .\nonumber \end{eqnarray}
Gluino masses are suppressed in this model by two factors of $(M/M_{GUT})$. 
One factor comes from the ratio of the R symmetry breaking scale and the 
messenger scale for color interactions, and the other from the ratio of the 
two different effective SUSY breaking scales, $\Lambda_C/\Lambda = M/M_{GUT}$.

The scalar masses squared arise at two-loops
 \begin{eqnarray} \tilde m_{squark}^2 = & 2 \Lambda_C^2 \left[
C_3\left({\alpha_3(M_{GUT}) \over 4 \pi}\right)^2\right] & \\
& + 2 \Lambda^2 \left[ C_2\left({\alpha_2(M)\over 4 \pi}\right)^2 
+ C_1 \left({\alpha_1(M)\over 4 \pi}\right)^2\right] &
 \nonumber \\   \tilde m_{slepton}^2 = m_{higgs}^2 = & 2 \Lambda^2 \left[
C_2\left({\alpha_2(M)\over 4 \pi}\right)^2 
+C_1 \left({\alpha_1(M)\over 4 \pi}\right)^2\right] &
 .\nonumber \end{eqnarray}
For $\Lambda_C/\Lambda = M/M_{GUT} << 1$, electroweak doublet squarks and 
sleptons are approximately degenerate while right-handed squarks and sleptons 
have mass ratios determined by the square of their weak hypercharge.  

\subsection{The gluino as the LSP}

The gluino-photino mass ratio, in our example, is given by 
\begin{equation} {m_{\tilde g} \over m_{\tilde \gamma}} \approx 
{3 \alpha_3\over 8 \alpha_{EM}} \left({M \over M_{GUT}}\right)^2  \end{equation}
The gluino is likely to be the lightest SUSY particle 
[LSP].\footnote{The gluino-gravitino mass ratio, given by ${ \alpha_3(M_Z) 
\sqrt{3}\over 4 \pi}{M M_{Planck} \over  M_{GUT}^2}{F_X \over F}$, is likely to 
be less than 1.} If the gluino is light and long-lived it would have some 
interesting new signatures for
detectors\cite{dawson,haber,farrar,hewett,gluinos}.   Several authors have
previously considered the possibility of gluinos being the  LSP\cite{farrar2}.  
Assuming R parity is conserved the gluino would be  absolutely stable.  It 
would form color singlet bound states  -- a pseudo-scalar gluino-gluino 
composite $\eta_{\tilde g}$ (called a gluinoball) and a fermionic gluino-gluon
composite $R_0$\cite{haber,farrar} (called a glueballino). For massless gluinos,
the $\eta_{\tilde g}$, $R_0$ and the scalar $0^{++}$ glueball form a massive 
supermultiplet with a dynamical mass determined by QCD.  For non-zero gluino 
mass, we have $m_{\eta_{\tilde g}} > m_{R_0} > m_{glueball}$.  In either case
$\eta_{\tilde g}$ like the glueball is unstable, decaying predominantly into 
multi-pion final states, while $R_0$ may remain stable.     These are just a
couple of states in the large menagerie of R-hadrons\cite{ff,rbaryons,rmesons}. 
There are also R-baryons (gluebarinos) and R-mesons (gluemesinos).  Of the
R-baryons, the flavor singlet $S_0 = \tilde g u d s$ is expected to be the
lightest\cite{rbaryons}.  It may also be long-lived or possibly even stable if
$m_{S_0} < m_{R_0} + m_n$.  Of  the R-mesons, the isotriplet $\tilde \rho$,
including the charged $\tilde \rho^+ = \tilde g u \bar d$ and neutral
$\tilde \rho^0 = \tilde g (u \bar u - d \bar d)$,
is expected to be the lightest\cite{rmesons}.  It may be short-lived, decaying
via strong interactions to $R_0 + \pi$, or long-lived, decaying only via weak
interactions to $R_0 + \mu^+ + \nu_{\mu}$ or $R_0 + e^+ + \nu_e$.  It could
even be stable, if it were the LSP.  The glueballino ($R_0$) and the R-rho 
($\tilde \rho^0$) are candidates for the LSP.  
We do not know apriori which of these is the lightest.  For gluino masses
greater than 3 GeV, bag model calculations suggest that $m_{\tilde \rho} \le
m_{R_0}$\cite{rmesons}.   However the ratio $m_{\tilde \rho}/m_{R_0}$ depends on
a gluon self-energy parameter which is not well determined.  
Comparison of different bag model calculations of glueball and mixed glue-$q \bar
q$ states leads us to expect uncertainties on the order of 100 - 200
MeV\cite{bag,farrar3}.   We shall thus consider two alternatives -- either $R_0$
or $\tilde \rho^0$  as the LSP.  We shall sometimes refer to the candidate
R-hadron LSP generically as $\tilde R$.  We shall also assume that the $S_0$ is
unstable and decays with a lifetime $\sim 10^{-10}$ seconds,  of order the
$\Lambda^0$ baryon lifetime.

The most dramatic change in SUSY phenomenology with a stable gluino LSP is that
the missing energy signature for SUSY is now gone.  Recall that general 
gluino searches in accelerator experiments have used the missing energy 
signature to identify the production and subsequent cascade decay of the
gluino.   A stable gluino, however, avoids all of these constraints.  In
our model, gluinos produced in high energy accelerator experiments will
hadronize until the $\tilde R$ is formed.  Since the $\tilde R$ is strongly
interacting it will be stopped in the hadronic calorimeter and all its kinetic
energy will be visible.  As a result, most quoted gluino limits\cite{haber} now
disappear.  However, for very heavy $\tilde R$s a missing energy signal
re-emerges which can, in principle, be used to place an upper limit on the 
gluino mass.  

Stable particle searches now provide the best limits on the $\tilde R$ mass. 
These are of two types --
\begin{itemize} \item mass spectrometer searches for heavy isotopes of hydrogen
or oxygen, and \item accelerator based fixed target or collider experiments
looking for stable charged or neutral hadrons. \end{itemize}   The mass
spectrometer searches for heavy isotopes of hydrogen provide severe constraints
for a stable gluino LSP.  

There have been several mass spectrometer searches for heavy (Z = 1) isotopes of
hydrogen\cite{muller,smith82}, referred to as $X^+$.   These experiments find no 
evidence for $X^+$ at an abundance relative to hydrogen given by
\begin{equation}{n_{X^+} \over n_{H^+}} > 10^{-18} - 10^{-22}\end{equation} 
for $m_{X^+}$ in the range from $2 - 350 $ GeV\cite{muller} or
\begin{equation}{n_{X^+} \over n_{H^+}} > 2\times10^{-28}\end{equation} for
$1000 > m_{X^+} > 12$ GeV\cite{smith82}.  In order for these results to  be
relevant for gluinos, however, we must interpret $X^+$ as a bound state of
$\tilde R$ (i.e. $ R_0$ or $\tilde \rho^0$) with hydrogen, deuterium or other
light element, i.e. 
$X^+ = \tilde R H^+, \tilde R D^+$.   Of course, this can only rule out a stable
$\tilde R$ if two conditions are satisfied --
\begin{enumerate}
\item $\tilde R$ in fact binds to light elements, and
\item $\tilde R$ is present at the experimentally accessible abundances.
\end{enumerate}
We shall argue later that the expected abundance of $\tilde R$ is such
that these experiments should have seen something if $\tilde R$ binds to
hydrogen.  Thus let us consider the question of binding.    The strong
interactions of $R_0$ or $\tilde \rho^0$ with hydrogen differ.   An
$R_0$ has strong interactions but it is not expected to behave like a neutron. 
The  dominant contribution to the binding of deuterium comes from pion exchange, 
while $R_0$, containing no valence quarks, is unlikely to couple significantly
to pions, rhos, etc.  The dominant exchange is likely to be a glueball with mass
of order 1.5  GeV.  Thus the interaction range is short, about ${ 1 \over 10}$
fm, and  considering that the deuterium binding energy is already quite small,
it is  unlikely that the $R_0$ would bind to light nuclei. Hence it would not be
seen in experiments searching for heavy isotopes of Hydrogen.  The $\tilde
\rho^0$, on the other hand, probably does exchange pions and is thus more likely
to bind to hydrogen.\footnote{Isospin symmetry, however, forbids the direct
$\tilde \rho^0 \tilde \rho^0 \pi^0$ coupling.}  This may exclude the
$\tilde \rho^0$ (assuming it is the LSP) with mass greater than 2 GeV. Clearly
further studies are necessary to decide this important issue.

Consider the case of $\tilde \rho^0$ as the LSP.  Then it is likely that the
charged states will either be stable or very long-lived.  \begin{itemize}
\item If the charged states are stable then this case is already ruled out 
by several complimentary experiments.  Searches for heavy isotopes of hydrogen
limit the relative abundance $$n_{\tilde \rho^+}/n_{H^+} <  10^{-18} $$ for 
$m_{\tilde \rho^+} > 2 GeV$\cite{muller,smith82}.  In addition, accelerator 
experiments exclude a long-lived $\tilde \rho^+$ with mass
in the range $4 < m_{\tilde \rho^+} < 10 GeV$ and lifetime 
$\tau_{\tilde \rho^+} > 5 \times 10^{-8}$ sec.~\cite{cutts} or with
$1.9 <  m_{\tilde \rho^+} < 13.6 GeV$ and $\tau_{\tilde \rho^+} >  10^{-7}$
sec.~\cite{akers}.  Thus we conclude that the $\tilde \rho^+$ cannot be 
stable (LSP or otherwise).\footnote{We believe that it is extremely unlikely 
for a stable $\tilde \rho$ to survive in the narrow window between 1.5 GeV, 
associated with the
expected dynamical mass due to QCD, and 1.9 GeV from accelerator and mass
spectrometer limits.}
\item If the $\tilde \rho^+$ is unstable, then its lifetime is expected to
be of order the neutron lifetime $\sim 10^3$ sec., since the $\tilde \rho^+ -
\tilde \rho^0$ mass difference would only be a few MeV and the decay would be
via the weak interaction process
$\tilde \rho^+ \rightarrow \tilde \rho^0 + e^+ + \nu_e$.  
 In this case, a Fermilab collider experiment is significant~\cite{abe}. It
excludes $\tilde \rho^+$ with mass greater than $50$ GeV.  Thus a long-lived,
 unstable $\tilde \rho^+$ is only allowed in the mass range
$13.6 < m_{\tilde \rho^+} < 50$  GeV.  Moreover, since $\tilde \rho^0$ and 
$\tilde \rho^+$ are nearly degenerate,  we have  $13.6 < m_{\tilde \rho^0} <
50$ GeV.
\end{itemize} 
Note, if $\tilde \rho^0$ binds to hydrogen, then the mass spectrometer
limit combined with the above accelerator data\cite{cutts,akers} excludes the
case $\tilde \rho^0$ as the LSP.  

Let us next consider the possibility that the glueballino, $R_0$, 
is the LSP and that the lightest charged or neutral gluebarino or gluemesino
decays into $R_0$ with lifetime shorter than $\sim 10^{-8}$ seconds.
What are the experimental limits on such a stable glueballino?   The mass range
between 2  and 4.2 GeV is presently excluded for
$m_{R_0}$\cite{dawson,haber,farrar}.    This follows from the absence of a peak
in the  photon energy spectrum in the process  $\Upsilon
\rightarrow \gamma  \eta_{\tilde g}$ by CUSB~\cite{cusb} or in the process
$\Upsilon' \rightarrow \gamma \chi_b(1^3P_1)$ followed by $\chi_b(1^3P_1)
\rightarrow g \tilde g \tilde g$ by Argus~\cite{argus} or from a search for new
hadrons with lifetimes greater than $10^{-7}$ sec, using time-of-flight in a
590m long  neutral beam at FNAL by Gustafson et al.\cite{gustafson,dawson}. 
Recently, the so-called light gluino window has been severely constrained by the
analysis of $e^+ e^- \rightarrow 4$ jets using LEP data~\cite{csikor,aleph}. 
The present limit set by Aleph~\cite{aleph} is $m_{\tilde g} > 6.3 $
GeV.  Note, however that this bound is subject to theoretical uncertainties of
higher order QCD effects on the 4 jet angular distributions\cite{farrar4}.  In
addition one should compare QCD with or without the effects of gluinos.   This
has not been done in this case.  If we ignore, for the moment, the Aleph limit,
then the light gluino window is still a viable option with only the mass range 
$2 < m_{R_0} < 4.2$ GeV excluded.  If on the other hand, we accept the Aleph
bound, then the mass range $m_{R_0} > 6.3$ GeV is still allowed.  Of course,
this assumes that $R_0$ does not bind to hydrogen.

We now argue that if $\tilde R$ binds to hydrogen, it is excluded with mass 
$m_{\tilde R} > 2$ GeV by mass spectrometer searches for heavy
hydrogen~\cite{muller,smith82}.  Given that $\tilde R$  binds to hydrogen forming
an $X^+$, we must calculate the expected abundance $n_{X^+}/n_{H^+}$.  The 
cosmological abundance of $\tilde R$ is expected to be of 
order\cite{zeld,steigman,dover}
\begin{equation}  {n_{\tilde R} \over n_H} \approx 10^{-10} \left( {m_{\tilde R}
\over 1 GeV}\right) . \end{equation}  $\tilde R$s will then bind to nuclei; some
will be processed into light elements beginning at nucleosynthesis and this
process will accelerate to heavier elements during stellar evolution. Those
contained in heavy elements are likely to remain there today, since they are
protected from annihilation by the coulomb barrier of the host nucleus.  Those
contained in light elements can undergo further annihilations in stellar
interiors at sufficiently high temperatures (of order $10^{10}$ $^{\circ}$K
corresponding to an MeV mean thermal energy).  However the $X^+$s which have not
been reprocessed through stars will remain today. The fraction of premordial 
hydrogen on earth (that which has not been processed through stars) is a number 
of order 1\cite{steigman2};  hence the terrestrial abundance of $X^+$s is not 
expected to be significantly different than the cosmological abundance of 
$\tilde R$s.  As a
consequence, $\tilde R$s with mass greater than $2$ GeV and which bind to
hydrogen are excluded.   If the light gluino window has been closed by Aleph
and $\tilde R$ binds to hydrogen, we must conclude that the gluino LSP scenario
is excluded.

Even if $\tilde R$ does not bind to hydrogen, it is still likely to 
bind to heavy nuclei.  In this context, a search for heavy isotopes
of oxygen is significant\cite{middleton}.  No evidence was found for such a
heavy isotope (HO) with an abundance $n_{HO}/n_{O} > {\rm few}\times10^{-19} -
10^{-16}$  for $ 20 < m_{HO} < 54$ amu corresponding to 
$ 4 < m_{R_0} < 38 $ GeV.  Whether or not this constrains the theory however
requires a detailed calculation of the processing of $\tilde R$ into heavy
elements.  A lower bound on the number of $\tilde R$s on earth can be obtained
by considering the flux of gluinos incident on the earth due to their
production by cosmic rays in the upper atmosphere~\cite{voloshin}.  A crude
estimate of this effect, assuming an incident flux of gluinos, 
$10^{-6} - 10^{-8} cm^{-2} s^{-1}$,\footnote{This flux is obtained using the known flux of cosmic ray
protons and the data on charm production to estimate the gluino production
rate\cite{voloshin}.} incident on a 4 km column depth
of water during the $10^9$ year age of the earth, gives an abundance 
$n_{\tilde R}/n_H \sim 10^{-18} - 10^{-20}$.   
Clearly a more careful calculation is needed.

To summarize,  if gluinos are the LSP then they will form stable hadrons.
A stable charged R-hadron is ruled out by accelerator and mass spectrometer
experiments.  A neutral $R_0$ or $\tilde \rho^0$, is however not 
necessarily ruled out.  For the glueballino ($R_0$), we assumed that the higher 
mass, unstable charged states would have
short enough lifetimes, $\le 10^{-8}$s, to escape detection.  For the $\tilde
\rho^0$, on the other hand, we assumed that the  $\tilde \rho^+$ is sufficiently
long-lived that it would not have escaped detection in accelerator experiments
looking for stable charged hadrons.
If either binds to hydrogen,  forming a stable heavy isotope $X^+$, then the
combination of accelerator experiments and  mass spectrometer searches excludes
it with any mass.  If  they do not bind to hydrogen, then the best limit on the
glueballino mass, 
$m_{R_0} > 6.3 $ GeV, comes from Aleph\cite{aleph}~\footnote{If the higher
order QCD corrections to the 4 jet angular correlations significantly affect
the Aleph results then this constraint may disappear.  In this case
the light gluino window remains open.}  and on the $\tilde \rho^0$ mass, $50  >
m_{\tilde \rho^0} > 13.6 GeV$, comes from accelerator
experiments\cite{cutts,akers,abe}.   Searches for 
anomalously heavy isotopes of heavy elements may constrain this mass range, but 
detailed calculations of the present abundance of such isotopes are needed.

Finally, let us briefly consider the new signatures for SUSY at high energy
accelerators.   Squark decay now proceeds via $\tilde q  \rightarrow q + 
\tilde g$, so that squark-anti-squark production at an $e^+ e^-$ collider 
results in 4 jets in the final state\cite{farrar5}.   Such events have been
observed by the Aleph collaboration at LEP II\cite{aleph2}.  Photinos would
decay into 3 jets containing a quark-anti-quark pair and a gluino. Thus,
slepton-anti-slepton production at an $e^+ e^-$ collider would have 2 leptons
and up to 6 jets in the final state.   

At a $pp$ or $\bar p p$ collider, gluino production will be copious.  An
energetic  gluino is expected to produce an hadronic jet containing a
single $\tilde R$.  The $\tilde R$ will most likely be stopped in the
detector; consequently all ``kinetic" energy is visible in the detector.  This
statement has two significant consequences.  Firstly, since 
only the kinetic energy ($E - m_{\tilde R}$) is visible in the calorimeter,
a very heavy $\tilde R$ would still have a missing energy signal.  This
effect could be used to bound the gluino mass from above.    In addition, in
the case that $\tilde R = R_0$ one must take into account that the cross section
for $R_0$ - nucleon scattering is expected to be only of order 40
$(m_{\pi}/m_{glueball})^2$ mb, since it is predominantly due to glueball
exchange in the t-channel.  Thus it is suppressed compared to the typical
nucleon-nucleon total cross-section. This effect must be taken into account when
modeling ${dE \over dx}$ for $R_0$ in a detector.   

As a final note, it has been suggested that a stable glueballino might produce
the observed anomallous muon showers from the X-ray binary star 
Cygnus X-3\cite{cygnus}. Although this possibility has been questioned by 
several authors\cite{voloshin,cygnet}, it is not excluded for 
$m_{\tilde R} \le 10$ GeV.

\section{Conclusion}

We have discussed gauge-mediated SUSY breaking at intermediate scales.  In this 
case the effective SUSY breaking scale $\Lambda = F_X/M \sim 10 - 100 \; TeV$
but both the messenger scale $M$ and the SUSY breaking scale $\sqrt{F_X}$ are
much greater than $\Lambda$.    

In the case $M \sim M_{GUT}/10$ and invariant under 
the grand unified gauge symmetry, we have boundary conditions (at $M$) for
gauginos, squarks and sleptons which match those of minimal supergravity (or
CMSSM) SUSY breaking schemes with $m_0$ and $M_{1/2}$ related.   The
advantage of the gauge mediated scheme is that the problem of flavor changing
neutral currents can be mitigated or completely eliminated, depending on the
messenger scale.

In a particular example, we showed that the gluino can naturally be the LSP with
very interesting consequences for experiment.  In this example, there are two
effective gauge-mediated SUSY breaking scales $\Lambda = {F_X \over M}
\sim 10 - 100 \; TeV$ for color singlet states and $\Lambda_C = {F_X \over
M_{GUT}}$ for color non-singlet states. 

The stable gluino window may be quite large.  The crucial question is whether
the LSP binds to hydrogen.  If it does, then the window is closed by mass
spectrometer searches for heavy isotopes of hydrogen.  If it does
not, then there are two possibilities --  a $\tilde \rho^0$ LSP is allowed for  $
50  > m_{\tilde \rho^0} > 13.6 GeV$, while a glueballino LSP is allowed for $
m_{R_0}
\ge 6.3$ GeV.  In either case,  accelerator experiments looking for missing
energy would not have seen the LSP, since the LSP is stopped in the hadronic
calorimeter.  Note, the high mass gluino limit could in principle be constrained
by missing energy, since the rest mass would escape detection.

Finally, we did not discuss the $\mu$ problem in this paper.  However, in our 
second example the messenger and Higgs sectors mix and there does not appear to
be a symmetry which prevents the $\mu$ term from being generated radiatively
(once SUSY and SO(10) breaking is included).  However a more detailed model
which includes the SO(10) breaking sector would be needed before this question
can be addressed further.

\vspace{.2 in}
\noindent
{\bf \large Appendix 1 --- Soft SUSY breaking masses}
\bigskip

Soft SUSY breaking mass terms by definition preserve the property of SUSY to
eliminate all quadratic divergences\cite{dg,girardello}.  However they have an
additional property  which concerns us here in the context of the flavor 
changing neutral current problem.  Above the messenger scale these mass terms
 no longer scale as local dimension two (for scalars) or three (for gauginos)
operators.  In fact they scale as non-local operators; as a consequence the
effective mass terms are momentum dependent at high energies.  For example for
scalars below the messenger scale we have
$$\tilde m^2  \phi^*
\phi
$$ where the mass squared 
$$\tilde m^2 =  {F_X^2 \over M^2} $$ is a constant.  However for external 
momentum $p >> M$ the mass squared is momentum dependent 
$$  m^2(p) \sim {F_X^2 \over p^2} .$$  So in general we have
$$  m^2(p) \sim {F_X^2 \over M^2 + p^2} .$$   Similarly, for gauginos we have
$$ m_{\lambda_i}(p) \sim {F_X M_R \over p^2}  $$ for $p >> M$.  $M_R$ is the R
symmetry breaking scale.

How does this affect the FCNC problem?  In the effective theory below $M$,
soft SUSY breaking scalar masses ($\tilde m^2$) are 
degenerate in a  given charge sector.  Thus flavor changing neutral current
effects are suppressed.  However GUT physics can induce threshold corrections
which violate flavor symmetries\cite{hkr}.   If however $M << M_{GUT}$ these
corrections are suppressed and we have  
$$  m_{scalar}^2 = \tilde m^2 ( 1 +  O(\left({M \over M_{GUT}}\right)^2)) .$$

\vspace{.2 in}
\noindent
{\bf \large Appendix 2 --- A Natural Theory}
\bigskip

Consider the O'Raifeartaigh type theory with three SO(10) singlets, 
$X_i, \; i=1,2,3$ with arbitrary couplings to the 10 and to two additional 
gauge singlets $P^+, \; P^-$.  We have the superspace potential
$$W = (\lambda_{1i} X_i) M^2 + (\lambda_{2i} X_i) P^+ \; P^- + (\lambda_{3i} 
X_i) 10^2 $$ where $i = 1,2,3$ is summed and $\lambda_{ni}, \; n=1,2,3$ are 9
arbitrary Yukawa couplings.  It is then always possible to define 3 new gauge
singlet fields $X,\;Y,\;Z$ in terms of $X_i$, so that the superspace potential
has the form $$W = X \, M^2 + (a_1 \, X + b_1 \, Y) P^+ \;P^- + (a_2 \, X + b_2
\, Y + Z) 10^2 .$$  This theory has no supersymmetric ground state since
\begin{eqnarray} - F_X^* = & a_1 \, P^+\;P^- + a_2 10^2 + M^2 &  \\
                 - F_Y^* = & b_1 \, P^+\;P^- + b_2 10^2  & \nonumber \\
                 - F_Z^* = & 10^2  & \nonumber 
\end{eqnarray}

Consider the local minimum  
\begin{eqnarray} < 10 > = 0, &  < P^+ > = - < P^- > = \rho,&   \\
a_1 \, < X > + b_1 \, < Y > = 0, &
  & \nonumber
\end{eqnarray} with
$$ z \equiv < Z >  \;\;\;{\rm and} \;\;\;  y \equiv (-a_1 \, < Y > + b_1 \,
<X>)/(a_1^2 + b_1^2)^{1/2} $$  arbitrary.

 At the minimum we have
$$\rho^2 = \left({a_1 \over a_1^2 + b_1^2}\right) M^2 $$   with
\begin{eqnarray} < F_Z > = < F_{10} > = < F_{P^{\pm}} > = 0 ,& a_1 \, < F_X > + 
b_1 \, < F_Y > = 0  & \end{eqnarray}
$$  < F_y > \equiv  (-a_1 \, < F_Y > + b_1 \,
< F_X >)/(a_1^2 + b_1^2)^{1/2} =  b_1 M^2/(a_1^2 + b_1^2)^{1/2}   $$
 Thus supersymmetry is broken.   The vevs for $y, \, z$ will be determined once
radiative corrections and supergravity effects are included.
The linear combination  $$ z' \equiv {(a_2 b_1 - b_2 a_1)\over 
\sqrt{a_1^2 + b_1^2}} y + z $$  appears in the effective superspace potential 
$$W_{effective} = z' 10^2  $$  and thus is expected to get a large vev as 
discussed in the paper.  The orthogonal linear combination does not appear in 
the superspace potential.

 \vspace{.2 in}
 \noindent
{\bf \large Acknowledgments}

 \bigskip
This work is partially supported by DOE contract DOE/ER/01545-708.  
I would like to thank R. Boyd, G. Farrar, H. Goldberg, M. Sokolov and G. 
Steigman for interesting discussions and suggestions.

 \end{document}